\documentclass{article}     
\usepackage{palatino}
\usepackage{mathpazo}
\usepackage{a4}
\usepackage{vmargin}
\setpapersize{A4}
\setmarginsrb{2cm}{1.5cm}{2cm}{1.5cm}{5mm}{5mm}{5mm}{5mm}
\usepackage{amsthm}
\theoremstyle{definition}
\newtheorem{definition}{Definition}
\newtheorem{theorem}{Theorem}

\usepackage{url}
\usepackage{color}
\usepackage{graphicx}
\usepackage{amsfonts}
\usepackage{amsmath}
\usepackage{amssymb}
\usepackage[shortcuts]{extdash}
\usepackage{xspace}
\usepackage{balance}
\usepackage{subcaption}
\usepackage{alltt}
\usepackage{verbatim}
\newenvironment{snippet}{\verbatim}{\endverbatim}

\newcommand*{\sref}[2]{(\texttt{#1}: \texttt{#2})}

\definecolor{tracegray}{gray}{0.5}
\newcommand{\trace}[1]{\textcolor{tracegray}{#1 :: \ }}

\newcommand{\vbar}{\ensuremath{\mid }}
\newcommand{\ruleSpace}{\;\;\;\;\;\;}
\newcommand{\ruleSpaceHalf}{\;\;\;}

\newcommand{\bslash}{\char`\\}

\newcommand{\code}[1]{\texttt{#1}}
\newcommand{\kw}[1]{\textsf{\small{\bfseries#1}}}
\newcommand{\kwn}[1]{\textsf{\bfseries#1}}
\newcommand{\hole}{\ensuremath{\Box}}
\newcommand*{\join}{\sqcup}

\newcommand*{\sqleq}{\sqsubseteq}

\newcommand*{\bwdF}{\ensuremath{\mathsf{bwd}}}
\newcommand*{\fwdF}{\ensuremath{\mathsf{fwd}}}
\newcommand*{\bwdFt}[2]{\ensuremath{\mathsf{bwd}^{#1}_{#2}}}
\newcommand*{\fwdFt}[2]{\ensuremath{\mathsf{fwd}^{#1}_{#2}}}
\newcommand*{\prefix}[1]{{\downarrow}#1}
\newcommand{\True}{\code{true}}
\newcommand{\False}{\code{false}}
\newcommand{\truekw}{\kw{true}}
\newcommand{\falsekw}{\kw{false}}
\newcommand{\truenkw}{\kwn{true}}
\newcommand{\falsenkw}{\kwn{false}}

\newcommand{\seq}{\,;\,}

\newcommand{\ifKw}{\code{if}\xspace}
\newcommand{\skipKw}{\code{skip}\xspace}
\newcommand{\impIf}{\code{if}\;b\; \code{then} \;\{\; c_1\; \} \; \code{else} \;\{\; c_2\; \}}

\newcommand{\whileKw}{\code{while }}
\newcommand{\impWhile}{\code{while}\;b\; \code{do} \;\{\; c\; \}}

\newcommand{\tracewhiletrue}{\trace{\code{while}_{\truenkw}\; T_b\; \code{do} \;\{\; T_c\; \};\; T_w}}
\newcommand{\tracewhilefalse}{\trace{\code{while}_{\falsenkw}\; T_b}}
\newcommand{\traceiftrue}{\trace{\code{if}_{\truenkw}\; T_b\;\code{then} \;\{\; T_1\; \}}}
\newcommand{\traceiffalse}{\trace{\code{if}_{\falsenkw}\; T_b\;\code{else} \;\{\; T_2\; \}}}

\author{Jan Stolarek
\and 
James Cheney
}

\title{Verified Self-Explaining Computation}

\begin{document}
\maketitle

\begin{abstract}
Common programming tools, like compilers, debuggers, and IDEs, crucially rely on
the ability to analyse program code to reason about its behaviour and
properties.  There has been a great deal of work on verifying compilers and
static analyses, but far less on verifying dynamic analyses such as program
slicing.  Recently, a new mathematical framework for slicing was introduced in
which forward and backward slicing are dual in the sense that they constitute a
Galois connection.  This paper formalises forward and backward dynamic slicing
algorithms for a simple imperative programming language, and formally verifies
their duality using the Coq proof assistant.
\end{abstract}

\section{Introduction}
\label{sec:intro}

The aim of mathematical program construction is to proceed from (formal)
specifications to (correct) implementations.  For example, critical components
such as compilers, and various static analyses they perform, have been
investigated extensively in a formal setting~\cite{Leroy-BKSPF-2016}.  However,
we unfortunately do not yet live in a world where all programs are constructed
in this way; indeed, since some aspects of programming (e.g. exploratory data
analysis) appear resistant to \emph{a priori} specification, one could debate
whether such a world is even possible. In any case, today programs ``in the
wild'' are not always mathematically constructed.  What do we do then?

One answer is provided by a class of techniques aimed at \emph{explanation},
\emph{comprehension} or \emph{debugging}, often based on run-time monitoring,
and sometimes with a pragmatic or even ad hoc flavour.  In our view, the
mathematics of constructing well-founded (meta)programs for explanation are
wanting~\cite{cheney13pbf}.  For example, dynamic analyses such as \emph{program
  slicing} have many applications in comprehending and restructuring programs,
but their mathematical basis and construction are far less explored compared to
compiler verification~\cite{Binkley06,Blazy15}.

Dynamic program slicing is a runtime analysis that identifies fragments of a
program's input and source code -- known together as a \emph{program slice} --
that were relevant to producing a chosen fragment of the output (a \emph{slicing
  criterion})~\cite{KorLas88,Weiser81}.  Slicing has a very large literature,
and there are a wide variety of dynamic slicing algorithms.  Most work on
slicing has focused on imperative or object-oriented programs.

One common application of dynamic slicing is program debugging.  Assume we have
a program with variables \code{x}, \code{y}, and \code{z} and a programmer
expects that after the program has finished running these variables will have
respective values \code{1}, \code{2}, and \code{3}.  If a programmer unfamiliar
with the program finds that after execution, variable \code{y} contains \code{1}
where she was expecting another value, she may designate \code{y} as a slicing
criterion, and dynamic slicing will highlight fragments of the source code that
could have contributed to producing the incorrect result.  This narrows down the
amount of code that the programmer needs to inspect to correct a program.  In
this tiny example, of course, there is not much to throw away and the programmer
can just inspect the program --- the real benefit of slicing is for
understanding larger programs with multiple authors.  Slicing can also be used
for program comprehension, i.e. to understand the behaviour of an already
existing program in order to re-engineer its specification, possibly
non-existent or not up-to-date.

In recent years a new semantic basis for dynamic slicing has been
proposed~\cite{perera12icfp,RicStoPerChe17}.  It is based on the concept of
Galois connections as known from order and lattice theory.  Given lattices $X$
and $Y$, a Galois connection is a pair of functions $g : Y \to X$ and $f : X \to
Y$ such that $g(y) \leq x \iff y \leq f(x)$; then $g$ is the \emph{lower
  adjoint} and $f$ is the \emph{upper adjoint}.  Galois connections have been
advocated as a basis for mathematical program construction already, for example
by Backhouse~\cite{backhouse00acmpc} and Mu and Oliveira~\cite{mu12jlamp}.  They
showed that if one can specify a problem space and show that it is one of the
component functions of a Galois connection (the ``easy'' part), then
\emph{optimal solutions} to the problem (the ``hard'' part) are uniquely
determined by the dual adjoint.  A simple example arises from the duality
between integer multiplication and division: the Galois connection $x \cdot y
\leq z \iff x \leq z/y$ expresses that $z/y$ is the greatest integer such that
$(z/y) \cdot y \leq z$.

Whereas Galois connections have been used previously for constructing programs
(as well as other applications such as program analysis), here we consider using
Galois connections to construct programs for program slicing.  In our setting,
we consider lattices of \emph{partial inputs} and \emph{partial outputs} of a
computation corresponding to possible input and output slicing criteria, as well
as \emph{partial programs} corresponding to possible slices---these are regarded
as part of the input.  We then define a forward semantics (corresponding to
\emph{forward slicing}) that expresses how much of the output of a program can
be computed from a given partial input.  Provided the forward semantics is
monotone and preserves greatest lower bounds, it is the upper adjoint of a
Galois connection, whose lower adjoint computes for each partial output the
\emph{smallest} partial input needed to compute it --- which we consider an
\emph{explanation}.  In other words, forward and backward slicing are
\emph{dual} in the sense that forward slicing computes ``as much as possible''
of the output given a partial input, while backward slicing computes ``as little
as needed'' of the input to recover a partial output.

\begin{figure}[tb]
\centering
\includegraphics[scale=0.3]{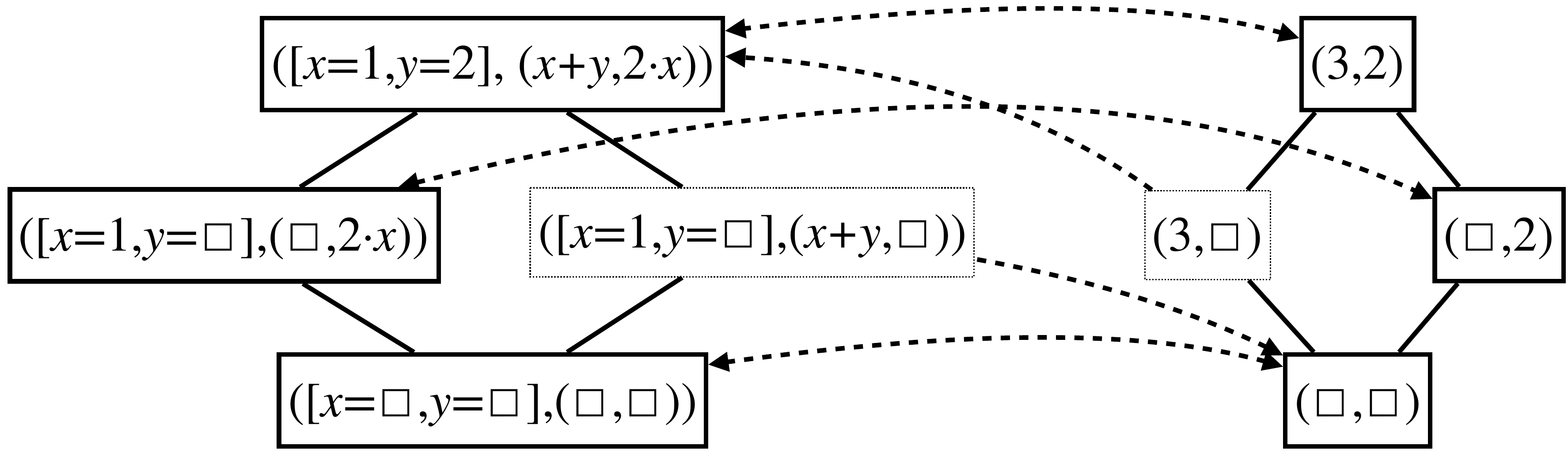}
\caption{Input and output lattices and Galois connection corresponding to
  expression $(x+y,2 \cdot x)$ evaluated with input $[x=1, y=2]$ and output
  $(3,2)$.  Dotted lines with arrows pointing left calculate the lower adjoint,
  those pointing right calculate the upper adjoint, and lines with both arrows
  correspond to the induced isomorphism between the minimal inputs and maximal
  outputs.  Several additional elements of the input lattice are
  omitted.}\label{fig:example}
\end{figure}

Figure~\ref{fig:example} illustrates the idea for a small example where the
``program'' is an expression $(x+y,2x)$, the input is an initial store
containing $[x=1,y=2]$ and the output is the pair $(3,2)$.  Partial inputs,
outputs, and programs are obtained by replacing subexpressions by ``holes''
($\Box$), as illustrated via (partial) lattice diagrams to the left and right.
In the forward direction, computing the first component succeeds if both $x$ and
$y$ are available while the second component succeeds if $x$ is available; the
backward direction computes the least input and program slice required for each
partial output.  Any Galois connection induces two isomorphic sublattices of the
input and output, and in Figure~\ref{fig:example} the elements of these
sublattices are enclosed in boxes with thicker lines.  In the input, these
elements correspond to \emph{minimal} explanations: partial inputs and slices in
which every part is needed to explain the corresponding output.  The
corresponding outputs are \emph{maximal} in a sense that their minimal
explanations do not explain anything else in the output.

The Galois connection approach to slicing has been originally developed for
functional programming languages~\cite{perera12icfp} and then extended to
functional languages with imperative features~\cite{RicStoPerChe17}.  So far it
has not been applied to conventional imperative languages, so it is hard to
compare directly with conventional slicing techniques.  Also, the properties of
the Galois connection framework in~\cite{perera12icfp,RicStoPerChe17} have only
been studied in pen-and-paper fashion.  Such proofs are notoriously tricky in
general and these are no exception; therefore, fully validating the metatheory
of slicing based on Galois connections appears to be an open problem.

In this paper we present forward and backward slicing algorithms for a simple
imperative language Imp and formally verify the correctness of these algorithms
in Coq.  Although Imp seems like a small and simple language, there are
nontrivial technical challenges associated with slicing in the presence of
mutable state, so our formalisation provides strong assurance of the correctness
of our solution.  To the best of our knowledge, this paper presents the first
formalisation of a Galois connection slicing algorithm for an imperative
language.  Compared with Ricciotti et al.~\cite{RicStoPerChe17}, Imp is a much
simpler language than they consider, but their results are not formalised;
compared with L\'echenet et al.~\cite{LecKosLeGt18}, we formalise dynamic rather
than static slicing.

In Section~\ref{sec:overview} we begin by reviewing the syntax of Imp, giving
illustrative examples of slicing, and then reviewing the theory of slicing using
the Galois connection framework, including the properties of minimality and
consistency. In Section~\ref{sec:theory} we introduce an instrumented, tracing
semantics for Imp and present the forward and backward slicing algorithms.  We
formalise all of the theory from Section~\ref{sec:theory} in Coq and prove their
duality.  Section~\ref{sec:formalisation} highlights key elements of our Coq
development, with full code available online~\cite{the-code}.
Section~\ref{sec:related-work} provides pointers to other related work.

\section{Overview}
\label{sec:overview}
\subsection{Imp slicing by example}
\label{sec:slicing-example}

\begin{figure}[ht]
\small
 \begin{tabular}{lrcl}
   Arithmetic expressions & $a$   & $::=$   & $n \vbar x \vbar a_1 + a_2$ \\ 
   Boolean expressions    & $b$   & $::=$   & $\True \vbar \False
                                              \vbar a_1 = a_2 \vbar  \neg b \vbar b_1 \wedge b_2 $ \\
   Imperative commands    & $c$   & $::=$   & $\skipKw \vbar x := a \vbar c_1 \seq c_2$ \\
                          &       & $\vbar$ & $\impWhile \vbar \code{if}\;b\; \code{then} \;\{\; c_1\; \}\;\code{else} \;\{\; c_2\; \}$\smallskip\\
   Values                 & $v$   & $::=$   & $v_a \vbar v_b$ \\
   State                  & $\mu$ & $::=$   & $\varnothing \vbar \mu, x \mapsto v_a$ ($x$ fresh)
 \end{tabular}
 \caption{Imp syntax}
 \label{fig:imp-syntax}
\end{figure}

For the purpose of our analysis we use a minimal imperative programming language
Imp used in some textbooks on programming languages,
e.g.~\cite{Nielson99,SF,Winskel}\footnote{In the literature Imp is also referred
  to as WHILE.}.  Imp contains arithmetic and logical expressions, mutable state
(a list of mappings from variable names to numeric values) and a small core of
imperative commands: empty instruction (\skipKw), variable assignments,
instruction sequencing, conditional \ifKw instructions, and \whileKw loops.  An
Imp program is a series of commands combined using a sequencing operator.  Imp
lacks more advanced features, such as functions or pointers.

Figure~\ref{fig:imp-syntax} shows Imp syntax.  We use letter $n$ to denote
natural number constants and $x$ to denote program variables.  In this
presentation we have omitted subtraction ($-$) and multiplication ($\cdot$) from
the list of arithmetic operators, and less-or-equal comparison ($\leq$) from the
list of boolean operators.  All these operators are present in our Coq
development and we omit them here only to make the presentation in the paper
more compact.  Otherwise the treatment of subtraction and multiplication is
analogous to the treatment of addition, and treatment of $\leq$ is analogous to
$=$.

Dynamic slicing is a program simplification technique that determines which
parts of the source code contributed to a particular program output.  For
example, a programmer might write this simple program in Imp:

\small
\begin{alltt}
if (y = 1) then \{ y := x + 1 \}
           else \{ y := y + 1 \} ;
z := z + 1
\end{alltt}
\normalsize
\noindent
and after running it with an input state $[ \code{x} \mapsto 1, \code{y} \mapsto
  0, \code{z} \mapsto 2 ]$ might (wrongly) expect to obtain output state $[
  \code{x} \mapsto 1, \code{y} \mapsto 2, \code{z} \mapsto 3 ]$.  However, after
running the program the actual output state will be $[ \code{x} \mapsto 1,
  \code{y} \mapsto 1, \code{z} \mapsto 3 ]$ with value of \code{y} differing
from the expectation.

We can use dynamic slicing to debug this program by asking for an explanation
which parts of the source code and the initial input state contributed to
incorrect output value of \code{y}.  We do this by formulating a \emph{slicing
  criterion}, where we replace all values that we consider irrelevant in the
output state (i.e. don't need an explanation for them) with \emph{holes}
(denoted with \hole):
\begin{displaymath}
[ \code{x} \mapsto \hole, \code{y} \mapsto 1, \code{z} \mapsto \hole ]
\end{displaymath}
\noindent
and a slicing algorithm might produce a program slice:

\small
\begin{alltt}
if (y = 1) then \{ \hole \}
           else \{ y := y + 1 \} ; \hole
\end{alltt}
\normalsize
\noindent
with a sliced input state $[ \code{x} \mapsto \hole, \code{y} \mapsto 0,
  \code{z} \mapsto \hole ]$.  This result indicates which parts of the original
source code and input state could be ignored when looking for a fault (indicated
by replacing them with holes), and which ones were relevant in producing the
output indicated in the slicing criterion.  The result of slicing narrows down
the amount of code a programmer has to inspect to locate a bug.  Here we can see
that only the input variable \code{y} was relevant in producing the result;
\code{x} and \code{z} are replaced with a \hole\ in the input state, indicating
their irrelevance.  We can also see that the first branch of the conditional was
not taken (unexpectedly!)  and that in the second branch \code{y} was
incremented to become $1$.  With irrelevant parts of the program hidden away it
is now easier to spot that the problem comes from a mistake in the initial
state.  The initial value of \code{y} should be changed to $1$ so that the first
branch of the conditional is taken and then \code{y} obtains an output value of
$2$ as expected.

Consider the same example, but a different output slicing criterion $[\code{x}
  \mapsto \hole, \code{y} \mapsto \hole, \code{z} \mapsto 3]$.  In this case, a
correctly sliced program is as follows:

\small
\begin{alltt}
\hole ; z := z + 1
\end{alltt}
\normalsize
\noindent
with the corresponding sliced input $[\code{x} \mapsto \hole, \code{y} \mapsto
  \hole, \code{z} \mapsto 2]$, illustrating that the entire first conditional
statement is irrelevant.  However, while the conclusion seems intuitively
obvious, actually calculating this correctly takes some work: we need to ensure
that none of the assignments inside the taken conditional branch affected
\code{z}, and conclude from this that the value of the conditional test \code{y
  = 1} is also irrelevant to the final value of \code{z}.

\subsection{A Galois connection approach to program slicing}
\label{sec:slicing}

Example in Section~\ref{sec:slicing-example} relies on an intuitive feel of how
backward slicing should behave.  We now address the question of how to make that
intuition precise and show that slicing using the Galois connection framework
introduced by Perera et al.~\cite{perera12icfp} offers an answer.

Consider these two extreme cases of backward slicing behaviour:

\begin{enumerate}
 \item for any slicing criterion backward slicing always returns a full program
   with no holes inserted;
 \item for any slicing criterion backward slicing always returns a \hole,
   i.e. it discards all the program code.
\end{enumerate}

\noindent
Neither of these two specifications is practically useful since they don't
fulfil our intuitive expectation of ``discarding program fragments irrelevant to
producing a given fragment of program output''.  The first specification does
not discard anything, bringing us no closer to understanding which code
fragments are irrelevant.  The second specification discards everything,
including the code necessary to produce the output we want to understand.  We
thus want a backward slicing algorithm to have two properties:

\begin{itemize}
 \item \textbf{consistency}: backward slicing retains code required to produce
   output we are interested in.
 \item \textbf{minimality}: backward slicing produces the smallest partial
   program and partial input state that suffice to achieve consistency.
\end{itemize}

\noindent
Our first specification above does not have the minimality property; the second
one does not have the consistency property. To achieve these properties we turn
to order and lattice theory.

We begin with defining \emph{partial Imp programs}
(Figure~\ref{fig:partial-imp-syntax}) by extending Imp syntax presented in
Figure~\ref{fig:imp-syntax} with \emph{holes} (denoted using \hole\ in the
semantic rules). A hole can appear in place of any arithmetic expression,
boolean expression, or command.  In the same way we allow values stored inside a
state to be mapped to holes.  For example:
\begin{displaymath}
\mu = [ \code{x} \mapsto 1, \code{y} \mapsto \Box ]
\end{displaymath}
\noindent
is a \emph{partial state} that maps variable \code{x} to $1$ and variable
\code{y} to a hole.  We also introduce operation $\varnothing_\mu$ that takes a
state $\mu$ and creates a partial state with the same domain as $\mu$ but all
variables mapped to \hole.  For example if $\mu = [ \code{x} \mapsto 1, \code{y}
  \mapsto 2 ]$ then $\varnothing_\mu = [ \code{x} \mapsto \Box, \code{y} \mapsto
  \Box ]$.  A partial state that maps all its variables to holes is referred to
as an \emph{empty partial state}.

\begin{figure}[t]
  \small
 \begin{tabular}{lrcl}
   Partial arithmetic expr. & $a$   & $::=$ & $\dots \vbar \Box$\\
   Partial boolean expr.    & $b$   & $::=$ & $\dots \vbar \Box$\\
   Partial commands    & $c$   & $::=$ & $\dots \vbar \Box$\\
   Partial state                  & $\mu$ &$::=$ & $\varnothing \vbar \mu, x \mapsto v_a \vbar \mu, x \mapsto \Box$ \\
 \end{tabular}
 \caption{Partial Imp syntax.  All elements of syntax from
   Figure~\ref{fig:imp-syntax} remain unchanged, only \hole\ are added.}
 \label{fig:partial-imp-syntax}
\end{figure}

Having extended Imp syntax with holes, we define partial ordering relations on
partial programs and partial states that consider holes to be syntactically
smaller than any other subexpression.  Figure~\ref{fig:partial-order-aexp} shows
the partial ordering relation for arithmetic expressions.  Definitions for
ordering of partial boolean expressions and partial commands are analogous.
Ordering for partial states is defined element-wise, thus requiring that two
states in the ordering relation have identical domains, i.e. store the same
variables in the same order.

\begin{figure}[t]
 \begin{displaymath}
  \dfrac{}
        {\hole \sqleq a}\ruleSpace
  \dfrac{}
        {n \sqleq n}\ruleSpace
  \dfrac{}
        {x \sqleq x}\ruleSpace
  \dfrac{a_1 \sqleq a_1'\ruleSpace a_2 \sqleq a_2'}
        {a_1 + a_2 \sqleq a_1' + a_2'}
 \end{displaymath}
 \caption{Ordering relation for partial arithmetic expressions.}
 \label{fig:partial-order-aexp}
\end{figure}

For every Imp program $p$, a set of all partial programs smaller than $p$ forms
a complete finite lattice, written $\prefix{p}$ with $p$ being the top and
\hole\ the bottom element of this lattice.  Partial states, arithmetic
expressions, and boolean expressions form lattices in the same way.  Moreover, a
pair of lattices forms a (product) lattice, with the ordering relation defined
component-wise: \begin{figure}[t]
 \begin{displaymath}
 a \join \hole = a
 \ruleSpace
 \hole \join a = a
 \ruleSpace
 n \join n = n
 \ruleSpace
 x \join x = x
 \end{displaymath}
 \begin{displaymath}
 (a_1 + a_2) \join (a_1' + a_2') = (a_1 \join a_1') + (a_2 \join a_2')
 \end{displaymath}
 \caption{Join operation for arithmetic expressions.}
 \label{fig:join-aexp}
\end{figure}

\begin{equation*}
\label{eqn:pair-lattice}
(a_1, b_1) \sqleq (a_2, b_2) \iff a_1 \sqleq a_2 \wedge b_1 \sqleq b_2
\end{equation*}

Figure~\ref{fig:join-aexp} shows definition of the \emph{join} (\emph{least
  upper bound}, $\join$) operation for arithmetic expressions.  Definitions for
boolean expressions and imperative commands are analogous.  A join exists for
every two elements from a complete lattice formed by a program $p$ or state
$\mu$ \cite[Theorem 2.31]{Davey}.

Assume we have a program $p$ paired with an input state $\mu$ that evaluates to
an output state $\mu'$.  We can now formulate slicing as a pair of functions
between lattices:

\begin{itemize}
 \item \textbf{forward slicing}: Forward slicing can be thought of as evaluation
   of partial programs.  A function \fwdFt{}{(p,\mu)} takes as its input a
   partial program and a partial state from a lattice formed by pairing a
   program $p$ and state $\mu$.  \fwdFt{}{(p,\mu)} outputs a partial state
   belonging to a lattice formed by $\mu'$.  The input to the forward slicing
   function is referred to as a \emph{forward slicing criterion} and output as a
   \emph{forward slice}.

 \item \textbf{backward slicing}: Backward slicing can be thought of as
   ``rewinding'' a program's execution.  A function \bwdFt{}{\mu'} takes as its
   input a partial state from the lattice formed by the output state $\mu'$.
   \bwdFt{}{\mu'} outputs a pair consisting of a partial program and a partial
   state, both belonging to a lattice formed by program $p$ and state $\mu$.
   Input to a backward slicing function is referred to as a \emph{backward
     slicing criterion} and output as a \emph{backward slice}.
\end{itemize}

A key point above (discussed in detail elsewhere~\cite{RicStoPerChe17}) is that
for imperative programs, both $\fwdFt{}{(p,\mu)}$ and $\bwdFt{}{\mu'}$ depend
not only on $p,\mu,\mu'$ but also on the particular execution path taken while
evaluating $p$ on $\mu$.  (In earlier work on slicing pure functional
programs~\cite{perera12icfp}, traces are helpful for \emph{implementing} slicing
efficiently but not required for \emph{defining} it.)  We make this information
explicit in Section~\ref{sec:theory} by introducing \emph{traces} $T$ that
capture the choices made during execution.  We will define the slicing
algorithms inductively as relations indexed by $T$, but in our Coq formalisation
\fwdFt{T}{(p,\mu)} and \bwdFt{T}{\mu'} are represented as dependently-typed
functions where $T$ is a proof term witnessing an operational derivation.

A pair of forward and backward slicing functions is guaranteed to have both the
minimality and consistency properties when they form a Galois
connection~\cite[Lemmas 7.26 and 7.33]{Davey}.

\begin{definition}[Galois connection]
\label{def:galois-connection}
Given lattices $P$, $Q$ and two functions $f : P \rightarrow Q$, $g : Q
\rightarrow P$, we say $f$ and $g$ form a Galois connection (written $f \dashv
g$) when $\forall_{p \in P, q \in Q} f(p) \sqsubseteq_Q q \iff p \sqsubseteq_P
g(q)$. We call $f$ a \emph{lower adjoint} and $g$ an \emph{upper adjoint}.
\end{definition}

\noindent
Importantly, for a given Galois connection $f \dashv g$, function $f$ uniquely
determines $g$ and vice versa~\cite[Lemma 7.33]{Davey}.  This means that our
choice of \fwdF\ (i.e. definition of how to evaluate partial programs on partial
inputs) uniquely determines the backward slicing function \bwdF\ that will be
minimal and consistent with respect to \fwdF, provided we can show that
\fwdF\ and \bwdF\ form a Galois connection.  There are many strategies to show
that two functions $f : P \rightarrow Q$ and $g : Q \rightarrow P$ form a Galois
connection, or to show that $f$ or $g$ in isolation has an upper or respectively
lower adjoint.  One attractive approach is to show that $f$ preserves least
upper bounds, or dually that $g$ preserves greatest lower bounds (in either
case, monotonicity follows as well).  This approach is indeed attractive because
it allows us to analyse just one of $f$ or $g$ and know that its dual adjoint
exists, without even having to write it down.  Indeed, in previous studies of
Galois slicing~\cite{perera12icfp,RicStoPerChe17}, this characterisation was the
one used: \fwdF\ was shown to preserve greatest lower bounds to establish the
existence of its lower adjoint \bwdF, and then efficient versions of \bwdF\ were
defined and proved correct.

For our constructive formalisation, however, we really want to give computable
definitions for both \fwdF\ and \bwdF\ and prove they form a Galois connection,
so while preservation of greatest lower bounds by \fwdF\ is a useful design
constraint, proving it does not really save us any work.  Instead, we will use
the following equivalent characterisation of Galois connections~\cite[Lemma
  7.26]{Davey}:

\begin{enumerate}

 \item\label{prop:gc-req-1} $f$ and $g$ are monotone

 \item\label{prop:gc-req-2} \emph{deflation} property holds:
  \begin{equation*}
   \forall_{q \in Q}\;\, f(g(q)) \sqsubseteq_Q q
  \end{equation*}

 \item\label{prop:gc-req-3} \emph{inflation} property holds:
  \begin{equation*}
   \forall_{p \in P}\;\, p \sqsubseteq_P g(f(p))
  \end{equation*}

\end{enumerate}

We use this approach in our Coq mechanisation.  We will first prove a general
theorem that any pair of functions that fulfils properties
(\ref{prop:gc-req-1})--(\ref{prop:gc-req-3}) above forms a Galois connection.
We will then define forward and backward slicing functions for Imp programs and
prove that they are monotone, deflationary, and inflationary.  Once this is done
we will instantiate the general theorem with our specific definitions of forward
and backward slicing to arrive at the proof that our slicing functions form a
Galois connection.  This is the crucial correctness property that we aim to
prove.  We also prove that existence of a Galois connection between forward and
backward slicing functions implies consistency and minimality properties.  Note
that consistency is equivalent to the inflation property.

\section{Dynamic program slicing}
\label{sec:theory}

\subsection{Tracing semantics}
\label{sec:imp}

\begin{figure}[t]
 \small
 \begin{tabular}{lrcl}
   Arithmetic traces & $T_a$ & $::=$   & ${n} \vbar {x(v_a)} \vbar {T_{a1} + T_{a2}}$ \\
   Boolean traces    & $T_b$ & $::=$   & ${\True} \vbar {\False}
   \vbar {T_{a1} = T_{a2}} \vbar {\neg T_b}$  
                                         $\vbar$ ${T_{b1} \wedge T_{b2}} $ \\
   Command traces    & $T_c$ & $::=$   & ${\skipKw} \vbar {x := T_a} \vbar {T_1 \seq T_2}$ \\
                     &       & $\vbar$ & ${\code{if}_{\truenkw}\;
                                         T_b\;\code{then} \;\{\; T_1\;
                                         \}}$
                                         $\vbar$  ${\code{if}_{\falsenkw}\; T_b\;\code{else} \;\{\; T_2\; \}}$\\
                     &       & $\vbar$ & ${\code{while}_{\falsenkw}\; T_b}$
                     $\vbar$  ${\code{while}_{\truenkw}\; T_b\; \code{do} \;\{\; T_{c}\; \};\; T_{w}}$\\
 \end{tabular}
 \caption{Trace syntax}
 \label{fig:imp-trace-syntax}
\end{figure}

Following previous work~\cite{perera12icfp,RicStoPerChe17}, we employ a
\emph{tracing semantics} to define the slicing algorithms.  Since dynamic
slicing takes account of the actual execution path followed by a run of a
program, we represent the execution path taken using an explicit trace data
structure.  Traces are then traversed as part of both the forward and backward
slicing algorithms. That is, unlike tracing evaluation, forward and backward
slicing follow the structure of traces, rather than the program.  Note that we
are not really inventing anything new here: in our formalisation, the trace is
simply a \emph{proof term} witnessing the derivability of the operational
semantics judgement.  The syntax of traces is shown in
Figure~\ref{fig:imp-trace-syntax}.  The structure of traces follows the
structure of language syntax with the following exceptions:

\begin{itemize}

 \item the expression trace $x(v_a)$ records both the variable name
   $x$ and a value $v_a$ that was read from program state $\mu$;

 \item for conditional instructions, traces record which branch was actually
   taken.  When the \ifKw condition evaluates to \truekw\ we store traces of
   evaluating the condition and the \code{then} branch; if it evaluates to
   \falsekw\ we store traces of evaluating the condition and the \code{else}
   branch.

 \item for \code{while} loops, if the condition evaluates to
   \falsekw\ (i.e. loop body does not execute) we record only a trace for the
   condition.  If the condition evaluates to \truekw\ we record traces for the
   condition ($T_b$), a single execution of the loop body ($T_{c}$) and the
   remaining iterations of the loop ($T_{w}$).
\end{itemize}

\begin{figure}[h!]
\vspace{-0.25cm}
 \small
 \begin{displaymath}
  \dfrac{}
        {\mu, n \Rightarrow \trace{n} v_n}
  \ruleSpace
  \dfrac{\mu(x) = v_a}
        {\mu, x \Rightarrow \trace{x(v_a)} v_a}
\ruleSpace
  \dfrac{\mu, a_1 \Rightarrow \trace{T_1} v_1 \ruleSpace \mu, a_2 \Rightarrow \trace{T_2} v_2}
        {\mu, a_1 + a_2 \Rightarrow \trace{T_1 + T_2} v_1 +_\mathbb{N} v_2 }
 \end{displaymath}



 \caption{Imp arithmetic expressions evaluation}
 \label{fig:imp-aexp-eval}
\end{figure}

\begin{figure}[h!]
\vspace{-0.25cm}
 \small
 \begin{displaymath}
  \dfrac{}
        {\mu, \True \Rightarrow \trace{\True} \truenkw}\ruleSpace
  \dfrac{}
        {\mu, \False \Rightarrow \trace{\False} \falsenkw}
 \end{displaymath}

 \begin{displaymath}
  \dfrac{\mu, a_1 \Rightarrow \trace{T_1} v_1 \ruleSpace \mu, a_2 \Rightarrow \trace{T_2} v_2}
        {\mu, a_1 = a_2 \Rightarrow \trace{T_1 = T_2} v_1 =_\mathbb{B} v_2}
  \ruleSpaceHalf
  \dfrac{\mu, b \Rightarrow \trace{T} v}
        {\mu, \neg b \Rightarrow \trace{\neg T} \neg_\mathbb{B} v}
 \end{displaymath}

 \begin{displaymath}
  \dfrac{\mu, b_1 \Rightarrow \trace{T_1} v_1 \ruleSpace \mu, b_2 \Rightarrow \trace{T_2} v_2}
        {\mu, b_1 \wedge b_2 \Rightarrow \trace{T_1 \wedge T_2} v_1 \wedge_\mathbb{B} v_2}
 \end{displaymath}

 \caption{Imp boolean expressions evaluation}
 \label{fig:imp-bexp-eval}
\end{figure}

\begin{figure*}[t]
 \begin{minipage}[b]{\textwidth}
 \small
 \begin{displaymath}
  \dfrac{}
        {\mu, \skipKw \Rightarrow \trace{\skipKw} \mu}\ruleSpace
  \dfrac{\mu, a\Rightarrow_a \trace{T_a} v_a \ruleSpace}
        {\mu, x := a \Rightarrow \trace{x := T_a}
        \mu[ x \mapsto v_a ]}
 \end{displaymath}

 \begin{displaymath}
  \dfrac{\mu, c_1 \Rightarrow \trace{T_1} \mu' \ruleSpace
         \mu', c_2 \Rightarrow \trace{T_2} \mu'' }
        {\mu, c_1 \seq c_2 \Rightarrow \trace{T_1 \seq T_2} \mu''}
 \end{displaymath}

 \begin{displaymath}
  \dfrac{\mu , b \Rightarrow \trace{T_b} \truekw \ruleSpace
         \mu, c_1 \Rightarrow \trace{T_1} \mu' }
        {\mu, \impIf \Rightarrow \traceiftrue \mu'}
 \end{displaymath}

 \begin{displaymath}
  \dfrac{\mu , b \Rightarrow \trace{T_b} \falsekw \ruleSpace
         \mu, c_2 \Rightarrow \trace{T_2} \mu' }
        {\mu, \impIf \Rightarrow \traceiffalse \mu'}
 \end{displaymath}

 \begin{displaymath}
  \dfrac{\mu, b \Rightarrow \trace{T_b} \falsekw}
        {\mu , \impWhile \Rightarrow \tracewhilefalse \mu}
 \end{displaymath}

 \begin{displaymath}
  \dfrac{\mu, b \Rightarrow \trace{T_b} \truekw \ruleSpace
         \mu, c \Rightarrow \trace{T_c} \mu'    \ruleSpace
         \mu', \impWhile \Rightarrow \trace{T_w}\mu''}
        {\mu , \impWhile \Rightarrow \tracewhiletrue \mu''}
 \end{displaymath}
 \end{minipage}

 \caption{Imp command evaluation.}
 \label{fig:imp-cmd-eval}
\end{figure*}

Figures~\ref{fig:imp-aexp-eval}--\ref{fig:imp-cmd-eval} show evaluation rules
for arithmetic expressions, boolean expressions, and imperative commands,
respectively\footnote{We overload the $\Rightarrow$ notation to mean one of
  three evaluation relations.  It is always clear from the arguments which
  relation we are referring to.}.  Traces are written in grey colour and
separated with a double colon ($::$) from the evaluation result.  Arithmetic
expressions evaluate to numbers (denoted $v_a$).  Boolean expressions evaluate
to either \truekw\ or \falsekw\ (jointly denoted as $v_b$).  Operators with
$_\mathbb{N}$ or $_\mathbb{B}$ subscripts should be evaluated as mathematical
and logical operators respectively to arrive at an actual value; this is to
distinguish them from the language syntax.  Commands evaluate by side-effecting
on the input state, producing a new state as output
(Figure~\ref{fig:imp-cmd-eval}).  Only arithmetic values can be assigned to
variables and stored inside a state.  Assignments to variables absent from the
program state are treated as no-ops.  This means all variables that we want to
write and read must be included (initialised) in the initial program state.  We
explain reasons behind this decision later in Section~\ref{sec:state-impl}.

\subsection{Forward slicing}
\label{sec:imp-forward-slicing}

In this and the next section we present concrete definitions of forward and
backward slicing for Imp programs.  Readers may prefer to skip ahead to
Section~\ref{sec:slicing-extended-example} for an extended example of these
systems at work first.  Our slicing algorithms are based on the ideas first
presented in~\cite{RicStoPerChe17}.  Presentation in Section~\ref{sec:slicing}
views the slicing algorithms as computable functions and we will implement them
in code as such.  However for the purpose of writing down the formal definitions
of our algorithms we will use a relational notation.  It is more concise and
allows easier comparisons with previous work.

Figures~\ref{fig:imp-fwd-aexp-slicing}--\ref{fig:imp-fwd-cmd-slicing} present
forward slicing rules for the Imp language\footnote{We again overload $\nearrow$
  and $\searrow$ arrows in the notation to denote one of three forward/backward
  slicing relations.  This is important in the rules for boolean slicing, whose
  premises refer to the slicing relation for arithmetic expressions, and command
  slicing, whose premises refer to slicing relation for boolean expressions.}.
As mentioned in Section~\ref{sec:slicing}, forward slicing can be thought of as
evaluation of partial programs.  Thus the forward slicing relations $\nearrow$
take a partial program, a partial state, and an execution trace as an input and
return a partial value, either a partial number (for partial arithmetic
expressions), a partial boolean (for partial logical expressions) or a partial
state (for partial commands). For example, we can read $\trace{T} \mu, c
\nearrow \mu'$ as ``Given trace $T$, in partial environment $\mu$ the partial
command $c$ forward slices to partial output $\mu'$.''

\begin{figure}[t!]

 \begin{displaymath}
  \dfrac{}
        {\trace{T} \mu, \Box \nearrow \Box}
  \ruleSpace
  \dfrac{}
        {\trace{n} \mu, n \nearrow n}
  \ruleSpace
  \dfrac{}
        {\trace{x(v_a)} \mu, x \nearrow \mu(x)}
 \end{displaymath}

 \begin{displaymath}
  \dfrac{\trace{T_1} \mu, a_1 \nearrow \Box}
        {\trace{T_1 + T_2} \mu, a_1 + a_2 \nearrow \Box}
  \ruleSpace
  \dfrac{\trace{T_2} \mu, a_2 \nearrow \Box}
        {\trace{T_1 + T_2} \mu, a_1 + a_2 \nearrow \Box}
 \end{displaymath}

 \begin{displaymath}
  \dfrac{\trace{T_1} \mu, a_1 \nearrow v_1 \ruleSpace \trace{T_2} \mu, a_2 \nearrow v_2}
        {\trace{T_1 + T_2} \mu, a_1 + a_2 \nearrow v_1 +_\mathbb{N} v_2}
        \;\;\; v_1, v_2 \neq \Box
 \end{displaymath}





 \caption{Forward slicing rules for Imp arithmetic expressions.}
 \label{fig:imp-fwd-aexp-slicing}
\end{figure}

\begin{figure}[tb!]

 \begin{displaymath}
  \dfrac{}
        {\trace{T} \mu, \Box \nearrow \Box}
 \end{displaymath}

 \begin{displaymath}
  \dfrac{}
        {\trace{\True} \mu, \True \nearrow \truekw}
  \ruleSpace
  \dfrac{}
        {\trace{\False} \mu, \False \nearrow \falsekw}
 \end{displaymath}

 \begin{displaymath}
  \dfrac{\trace{T_1} a_1 \nearrow \Box}
        {\trace{T_1 = T_2} \mu, a_1 = a_2 \nearrow \Box}
  \ruleSpace
  \dfrac{\trace{T_2} a_2 \nearrow \Box}
        {\trace{T_1 = T_2} \mu, a_1 = a_2 \nearrow \Box}
 \end{displaymath}

 \begin{displaymath}
  \dfrac{\trace{T_1} a_1 \nearrow v_1 \ruleSpace \trace{T_2} a_2 \nearrow v_2}
        {\trace{T_1 = T_2} \mu, a_1 = a_2 \nearrow v_1 =_\mathbb{B} v_2}
        \;\;\; v_1, v_2 \neq \Box
 \end{displaymath}



 \begin{displaymath}
  \dfrac{\trace{T} b \nearrow \Box}
        {\trace{\neg T} \mu, \neg b \nearrow \Box}
  \ruleSpace
  \dfrac{\trace{T} b \nearrow v_b}
        {\trace{\neg T} \mu, \neg b \nearrow \neg_\mathbb{B} v}
        \;\;\; v_b \neq \Box
 \end{displaymath}

 \begin{displaymath}
  \dfrac{\trace{T_1} b_1 \nearrow \Box}
        {\trace{T_1 \wedge T_2} \mu, b_1 \wedge b_2 \nearrow \Box}
  \ruleSpace
  \dfrac{\trace{T_2} b_2 \nearrow \Box}
        {\trace{T_1 \wedge T_2} \mu, b_1 \wedge b_2 \nearrow \Box}
 \end{displaymath}

 \begin{displaymath}
  \dfrac{\trace{T_1} b_1 \nearrow v_1 \ruleSpace \trace{T_2} b_2 \nearrow v_2}
        {\trace{T_1 \wedge T_2} \mu, b_1 \wedge b_2 \nearrow v_1 \wedge_\mathbb{B} v_2}
        \;\;\; v_1, v_2 \neq \Box
 \end{displaymath}

 \caption{Forward slicing rules for Imp boolean expressions.}
 \label{fig:imp-fwd-bexp-slicing}
\end{figure}

\begin{figure*}[ht!]
 \begin{minipage}[b]{\textwidth}
 \small
 \begin{displaymath}
  \dfrac{}
        {\trace{\skipKw} \mu, \Box \nearrow \mu}
  \ruleSpace
  \dfrac{}
        {\trace{\skipKw} \mu, \skipKw \nearrow \mu}
  \ruleSpace
  \dfrac{}
        {\trace{x := T_a} \mu, \Box \nearrow \mu[ x \mapsto \Box]}
 \end{displaymath}

 \begin{displaymath}
  \dfrac{\trace{T_a} \mu,a \nearrow v_a}
        {\trace{x := T_a} \mu, x := a \nearrow \mu[ x \mapsto v_a]}
  \ruleSpace
  \dfrac{\trace{T_1} \mu, \Box \nearrow \mu' \ruleSpace
         \trace{T_2} \mu', \Box \nearrow \mu''}
        {\trace{T_1 \seq T_2} \mu, \Box \nearrow \mu''}
 \end{displaymath}

 \begin{displaymath}
  \dfrac{\trace{T_1} \mu , c_1 \nearrow \mu' \ruleSpace
         \trace{T_2} \mu', c_2 \nearrow \mu''}
        {\trace{T_1 \seq T_2} \mu, c_1 \seq c_2 \nearrow \mu''}
  \dfrac{\trace{T_1} \mu  ,\Box \nearrow \mu'}
        {\traceiftrue \mu, \Box \nearrow \mu'}
 \end{displaymath}

 \begin{displaymath}
  \dfrac{\trace{T_b} \mu, b    \nearrow \Box \ruleSpace
         \trace{T_1} \mu, \Box \nearrow \mu'}
        {\traceiftrue \mu, \impIf \nearrow \mu'}
 \end{displaymath}

 \begin{displaymath}
  \dfrac{\trace{T_b} \mu, b   \nearrow v_b \ruleSpace
         \trace{T_1} \mu, c_1 \nearrow \mu'}
        {\traceiftrue \mu, \impIf \nearrow \mu'}
        \;\;\; v_b \neq \Box
 \end{displaymath}

 \begin{displaymath}
  \dfrac{\trace{T_2} \mu, \Box \nearrow \mu'}
        {\traceiffalse \mu, \Box \nearrow \mu'}
 \end{displaymath}

 \begin{displaymath}
  \dfrac{\trace{T_b} \mu, b    \nearrow \Box \ruleSpace
         \trace{T_2} \mu, \Box \nearrow \mu'}
        {\traceiffalse \mu, \impIf \nearrow \mu'}
 \end{displaymath}

 \begin{displaymath}
  \dfrac{\trace{T_b} \mu, b   \nearrow v_b \ruleSpace
         \trace{T_2} \mu, c_2 \nearrow \mu'}
        {\traceiffalse \mu, \impIf \nearrow \mu'}
        \;\;\; v_b \neq \Box
 \end{displaymath}

 \begin{displaymath}
  \dfrac{}
        {\tracewhilefalse \mu, \Box \nearrow \mu}
  \ruleSpace
  \dfrac{}
        {\tracewhilefalse \mu, \impWhile \nearrow \mu}
 \end{displaymath}

 \begin{displaymath}
  \dfrac{\trace{T_c} \mu  ,\Box \nearrow \mu_c \ruleSpace
         \trace{T_w} \mu_c,\Box \nearrow \mu_w}
        {\tracewhiletrue \mu, \Box \nearrow \mu_w}
 \end{displaymath}

 \begin{displaymath}
  \dfrac{\trace{T_b} \mu,  b    \nearrow \Box   \ruleSpace
         \trace{T_c} \mu,  \Box \nearrow \mu_c  \ruleSpace
         \trace{T_w} \mu_c,\Box \nearrow \mu_w}
        {\tracewhiletrue \mu, \impWhile \nearrow \mu_w}
 \end{displaymath}

 \begin{displaymath}
  \dfrac{\trace{T_b} \mu,b \nearrow v_b    \ruleSpace
         \trace{T_c} \mu,c \nearrow \mu_c  \ruleSpace
         \trace{T_w} \mu_c,\impWhile \nearrow \mu_w}
        {\tracewhiletrue \mu, \impWhile \nearrow \mu_w} \;\;\; v_b \neq \Box
 \end{displaymath}
 \end{minipage}

 \caption{Forward slicing rules for Imp commands.}
 \label{fig:imp-fwd-cmd-slicing}
\end{figure*}

A general principle in the forward slicing rules for arithmetic expressions
(Figure~\ref{fig:imp-fwd-aexp-slicing}) and logical expressions
(Figure~\ref{fig:imp-fwd-bexp-slicing}) is that ``holes propagate''.  This means
that whenever \hole\ appears as an argument of an operator, application of that
operator forward slices to a \hole.  For example, $1 + \hole$ forward slices to
a \hole\ and so does $\neg\hole$.  In other words, if an arithmetic or logical
expression contains at least one hole it will reduce to a \hole; if it contains
no holes it will reduce to a proper value.  This is not the case for commands
though.  For example, command
$\code{if}\;\code{true}\;\code{then}\;1\;\code{else}\;\hole$ forward slices to
$1$, even though it contains a hole in the (not taken) \code{else} branch.

A rule worth attention is one for forward slicing of variable reads:
\begin{displaymath}
 \dfrac{}{\trace{x(v_a)} \mu, x \nearrow \mu(x)}
\end{displaymath}
\noindent
It is important here that we read the return value from $\mu$ and not $v_a$
recorded in a trace.  This is because $\mu$ is a partial state and also part of
a forward slicing criterion.  It might be that $\mu$ maps $x$ to \hole, in which
case we must forward slice to \hole\ and not to $v_a$.  Otherwise minimality
would not hold.

Forward slicing rules for arithmetic and logical expressions both have a
universal rule for forward slicing of holes that applies regardless of what the
exact trace value is:
\begin{displaymath}
 \dfrac{}{\trace{T} \mu, \Box \nearrow \Box}
\end{displaymath}
\noindent
There is no such rule for forward slicing of commands
(Figure~\ref{fig:imp-fwd-cmd-slicing}).  There we have separate rules for
forward slicing of holes for each possible trace.  This is due to the
side-effecting nature of commands, which can mutate the state through variable
assignments.  Consider this rule for forward slicing of assignments w.r.t. a
\hole\ as a slicing criterion\footnote{When some partial value $v$ is used as a
  slicing criterion we say that we ``slice w.r.t. $v$''.}:
\begin{displaymath}
 \dfrac{}{\trace{x := T_a} \mu, \Box \nearrow \mu[ x \mapsto \Box]}
\end{displaymath}
\noindent
When forward slicing an assignment w.r.t. a \hole\ we need to erase (i.e. change
to a \hole) variable $x$ in the state $\mu$, which follows the principle of
``propagating holes''.  Here having a trace is crucial to know which variable
was actually assigned during the execution.  Rules for forward slicing of other
commands w.r.t.  a \hole\ traverse the trace recursively to make sure that all
variable assignments within a trace are reached.  For example:
\begin{displaymath}
 \dfrac{\trace{T_1}  \mu, \Box \nearrow \mu'}
       {\traceiftrue \mu, \Box \nearrow \mu'}
\end{displaymath}
\noindent
In this rule, trace $T_1$ is traversed recursively to arrive at a state $\mu'$
that is then returned as the final product of the rule.  Notice how trace $T_b$
is \emph{not} traversed.  This is because boolean expressions (and arithmetic
ones as well) do not have side effects on the state and so there is no need to
traverse them.

The problem of traversing the trace recursively to handle side-effects to the
state can be approached differently.  Authors of~\cite{RicStoPerChe17} have
formulated a single rule, which we could adapt to our setting like this:
\begin{displaymath}
 \dfrac{\mathcal{L} = \mathsf{writes}(T)} {\trace{T} \mu, \Box \nearrow \mu
   \triangleleft \mathcal{L}}
\end{displaymath}
\noindent
In this rule $\mathsf{writes}(T)$ means ``all state locations written to inside
trace $T$'' and $\mu \triangleleft \mathcal{L}$ means erasing (i.e. mapping to a
\hole) all locations in $\mu$ that are mentioned in $\mathcal{L}$.  Semantically
this is equivalent to our rules -- both approaches achieve the same effect.
However, we have found having separate rules easier to formalise in a proof
assistant.

\subsection{Backward slicing}
\label{sec:imp-backward-slicing}

Backward slicing rules are given in
Figures~\ref{fig:imp-bwd-aexp-slicing}--\ref{fig:imp-bwd-cmd-slicing}.  These
judgements should be read left-to-right, for example, $\trace{T} \mu \searrow
\mu', c$ should be read as ``Given trace $T$ and partial output state $\mu$,
backward slicing yields partial input $\mu'$ and partial command $c$.''  Their
intent is to reconstruct the smallest program code and initial state that
suffice to produce, after forward slicing, a result that is at least as large as
the backward slicing criterion.  To this end, backward slicing crucially relies
on execution traces as part of input, since slicing effectively runs a program
backwards (from final result to source code).

Figures~\ref{fig:imp-bwd-aexp-slicing} and \ref{fig:imp-bwd-bexp-slicing} share
a universal rule for backward slicing w.r.t. a hole.
\begin{displaymath}
 \dfrac{}{\trace{T} \mu, \Box \searrow \varnothing_\mu, \Box}
\end{displaymath}
\noindent
This rule means that to obtain an empty result it always suffices to have an
empty state and no program code.  This rule always applies preferentially over
other rules, which means that whenever a value, such as $v_a$ or $v_b$, appears
as a backward slicing criterion we know it is not a \hole.  Similarly,
Figure~\ref{fig:imp-bwd-cmd-slicing} has a rule:
\begin{figure}[t!]

 \begin{displaymath}
  \dfrac{}
        {\trace{T} \mu, \Box \searrow \varnothing_\mu, \Box}
  \ruleSpace
  \dfrac{}
        {\trace{n} \mu, v_a \searrow \varnothing_\mu, n}
 \end{displaymath}

 \begin{displaymath}
  \dfrac{}
        {\trace{x(v_a)} \mu, v_a \searrow \varnothing_\mu[ x \mapsto v_a ], x}
 \end{displaymath}

 \begin{displaymath}
  \dfrac{\trace{T_1} \mu, v_1 \searrow \mu_1, a_1 \ruleSpace \trace{T_2} \mu, v_2 \searrow \mu_2, a_2}
        {\trace{T_1 + T_2} \mu, v_a \searrow \mu_1 \join \mu_2, a_1 + a_2 }
 \end{displaymath}



 \caption{Backward slicing rules for Imp arithmetic expressions.}
 \label{fig:imp-bwd-aexp-slicing}
\end{figure}
 \begin{figure}[t!]
 \small
 \begin{displaymath}
  \dfrac{}
        {\trace{T} \mu, \Box \searrow \varnothing_\mu, \Box}
 \end{displaymath}

 \begin{displaymath}
  \dfrac{}
        {\trace{\True} \mu, \truenkw \searrow \varnothing_\mu, \True}
  \ruleSpace
  \dfrac{}
        {\trace{\False} \mu, \falsenkw \searrow \varnothing_\mu, \False}
 \end{displaymath}

 \begin{displaymath}
  \dfrac{\trace{T_1} \mu, v_1 \searrow \mu_1, a_1 \ruleSpace \trace{T_2} \mu, v_2 \searrow \mu_2, a_2}
        {\trace{T_1 = T_2} \mu, v_b \searrow \mu_1 \join \mu_2, a_1 = a_2 }
 \end{displaymath}


 \begin{displaymath}
  \dfrac{\trace{T} \mu, v_b \searrow \mu', b}
        {\trace{\neg T} \mu, v_b \searrow \mu', \neg b }
 \end{displaymath}

 \begin{displaymath}
  \dfrac{\trace{T_1} \mu, v_1 \searrow \mu_1, b_1 \ruleSpace \trace{T_2} \mu, v_2 \searrow \mu_2, b_2}
        {\trace{T_1 \wedge T_2} \mu, v_b \searrow \mu_1 \join \mu_2, b_1 \wedge b_2 }
 \end{displaymath}

 \caption{Backward slicing rules for Imp boolean expressions.}
 \label{fig:imp-bwd-bexp-slicing}
\end{figure}

\begin{figure*}[ht!]
 \begin{minipage}[b]{\textwidth}
 \small
 \begin{displaymath}
  \dfrac{}{\trace{T} \varnothing \searrow \varnothing_\varnothing, \Box}
  \ruleSpace
  \dfrac{}
        {\trace{\skipKw} \mu \searrow \mu, \Box}
  \ruleSpace
  \dfrac{}
        {\trace{x := T_a} \mu[x \mapsto \Box] \searrow
         \mu[x \mapsto \Box], \Box}
 \end{displaymath}

 \begin{displaymath}
  \dfrac{\trace{T_a} v_a \searrow \mu_a, a}
        {\trace{x := T_a} \mu[x \mapsto v_a] \searrow
         \mu_a \join \mu[x \mapsto \Box], x := a}
        \;\;\; v_a \neq \Box
 \end{displaymath}

 \begin{displaymath}
  \dfrac{\trace{T_2} \mu \searrow \mu', \Box \ruleSpace
         \trace{T_1} \mu' \searrow \mu'', \Box}
        {\trace{T_1 \seq T_2} \mu \searrow \mu'', \Box}
  \ruleSpace
  \dfrac{\trace{T_2} \mu \searrow \mu', \Box \ruleSpace
         \trace{T_1} \mu' \searrow \mu'', c_1}
        {\trace{T_1 \seq T_2} \mu \searrow \mu'', c_1 \seq \Box}
        \;\;\; c_1 \neq \Box
 \end{displaymath}

 \begin{displaymath}
  \dfrac{\trace{T_2} \mu \searrow \mu', c_2 \ruleSpace
         \trace{T_1} \mu' \searrow \mu'', c_1}
        {\trace{T_1 \seq T_2} \mu \searrow \mu'', c_1 \seq c_2}
        \;\;\; c_2 \neq \Box
  \ruleSpace
  \dfrac{\trace{T_1} \mu \searrow \mu', \Box}
        {\traceiftrue \mu \searrow \mu', \Box}
 \end{displaymath}

 \begin{displaymath}
  \dfrac{\trace{T_1} \mu \searrow \mu', c_1 \ruleSpace
         \trace{T_b} \texttt{true} \searrow \mu_b, b}
        {\traceiftrue \mu \searrow \mu' \join \mu_b,
         \texttt{if}\;b\; \texttt{then} \;\{\; c_1\; \} \; \texttt{else}
          \;\{\; \Box \; \}}
        \;\;\; c_1 \neq \Box
 \end{displaymath}

 \begin{displaymath}
  \dfrac{\trace{T_2} \mu \searrow \mu', \Box}
        {\traceiffalse \mu \searrow \mu', \Box}
 \end{displaymath}

 \begin{displaymath}
  \dfrac{\trace{T_2} \mu \searrow \mu', c_2 \ruleSpace
         \trace{T_b} \texttt{false} \searrow \mu_b, b}
        {\traceiffalse \mu \searrow \mu' \join \mu_b,
         \texttt{if}\;b\; \texttt{then} \;\{\; \Box \; \} \; \texttt{else}
          \;\{\; c_2 \; \}}
        \;\;\; c_2 \neq \Box
 \end{displaymath}

 \begin{displaymath}
  \dfrac{}
        {\tracewhilefalse \mu \searrow \mu, \Box}
  \ruleSpace
  \dfrac{\trace{T_w} \mu  \searrow \mu_w , \Box \ruleSpace
         \trace{T_c} \mu_w \searrow \mu_c, \Box}
        {\tracewhiletrue \mu \searrow \mu_c, \Box}
 \end{displaymath}

 \begin{displaymath}
  \dfrac{\trace{T_w} \mu           \searrow \mu_w  , \Box  \ruleSpace
         \trace{T_c} \mu_w         \searrow \mu_c , c     \ruleSpace
         \trace{T_b} \texttt{true} \searrow \mu_b , b }
        {\tracewhiletrue \mu \searrow \mu_c \join \mu_b, \impWhile }
        \;\;\; c \neq \Box
 \end{displaymath}

 \begin{displaymath}
  \dfrac{\trace{T_w} \mu           \searrow \mu_w  , c_w  \ruleSpace
         \trace{T_c} \mu_w          \searrow \mu_c , c    \ruleSpace
         \trace{T_b} \texttt{true} \searrow \mu_b , b }
        {\tracewhiletrue \mu \searrow \mu_c \join \mu_b, c_w \join \impWhile }
        \;\;\; c_w \neq \Box
 \end{displaymath}
 \end{minipage}

 \caption{Backward slicing rules for Imp commands.}
 \label{fig:imp-bwd-cmd-slicing}
\end{figure*}

\begin{displaymath}
 \dfrac{}{\trace{T} \varnothing \searrow \varnothing_\varnothing, \Box}
\end{displaymath}
\noindent
It means that backward slicing w.r.t. a state with an empty domain
(i.e. containing no variables) returns an empty partial state and an empty
program.  Of course having a program operating over a state with no variables
would be completely useless -- since a state cannot be extended with new
variables during execution we wouldn't observe any effects of such a program.
However, in the Coq formalisation, it is necessary to handle this case because
otherwise Coq will not accept that the definition of backward slicing is a total
function.

In the rule for backward slicing of variable reads (third rule in
Figure~\ref{fig:imp-bwd-aexp-slicing}) it might seem that $v_a$ stored inside a
trace is redundant because we know what $v_a$ is from the slicing criterion.
This is a way of showing that variables can only be sliced w.r.t. values they
have evaluated to during execution.  So for example if $x$ evaluated to $17$ it
is not valid to backward slice it w.r.t. $42$.

The rule for backward slicing of addition in
Figure~\ref{fig:imp-bwd-aexp-slicing} might be a bit surprising.  Each of the
subexpressions is sliced w.r.t. a value that this expression has evaluated to
($v_1$, $v_2$), and not w.r.t. $v_a$.  It might seem we are getting $v_1$ and
$v_2$ out of thin air, since they are not directly recorded in a trace.  Note
however that knowing $T_1$ and $T_2$ allows to recover $v_1$ and $v_2$ at the
expense of additional computations.  In the actual implementation we perform
induction on the structure of evaluation derivations, which record values of
$v_1$ and $v_2$, thus allowing us to avoid extra computations.  We show $v_1$
and $v_2$ in our rules but avoid showing the evaluation relation as part of
slicing notation.  This is elaborated further in Section~\ref{sec:slicing-impl}.

Recursive backward slicing rules also rely crucially on the join ($\join$)
operation, which combines smaller slices from slicing subexpressions into one
slice for the whole expression.

There are two separate rules for backward slicing of variable assignments (rules
3 and 4 in Figure~\ref{fig:imp-bwd-cmd-slicing}).  If a variable is mapped to a
\hole\ it means it is irrelevant.  We therefore maintain mapping to a \hole\ and
do not reconstruct variable assignment instructions.  If a variable is relevant
though, i.e. it is mapped to a concrete value in a slicing criterion, we
reconstruct the assignment instruction together with an arithmetic expression in
the RHS.  We also join state $\mu_a$ required to evaluate the RHS with $\mu[
  \code{x} \mapsto \hole]$.  It is crucial that we erase $x$ in $\mu$ prior to
joining.  Firstly, if \code{x} is assigned, its value becomes irrelevant prior
to the assignment, unless \code{x} is read during evaluation of the RHS (e.g. we
are slicing an assignment \code{x := x + 1}).  In this case \code{x} will be
included in $\mu_a$ but its value can be different than the one in $\mu$.  It is
thus necessary to erase \code{x} in $\mu$ to make a join operation possible.

At this point, it may be helpful to review the forward rules for assignment and
compare with the backward rules, illustrated via a small example.  Suppose we
have an assignment \code{z := x + y}, initially evaluated on $[\code{w} \mapsto
  0, \code{x} \mapsto 1, \code{y} \mapsto 2, \code{z} \mapsto 42]$, and yielding
result state $[\code{w} \mapsto 0, \code{x} \mapsto 1, \code{y} \mapsto 2,
  \code{z} \mapsto 3]$.  The induced lattice of minimal inputs and maximal
outputs consists of the following pairs:
\begin{eqnarray*}
{}  [\code{w} \mapsto v, \code{x} \mapsto 1, \code{y} \mapsto 2, \code{z} \mapsto \Box] & \longleftrightarrow &
    [\code{w} \mapsto v, \code{x} \mapsto 1, \code{y} \mapsto 2, \code{z} \mapsto 3]\\
{}  [\code{w} \mapsto v, \code{x} \mapsto 1, \code{y} \mapsto \Box, \code{z} \mapsto \Box] & \longleftrightarrow &
    [\code{w} \mapsto v, \code{x} \mapsto 1, \code{y} \mapsto \Box, \code{z} \mapsto \Box]\\
{}  [\code{w} \mapsto v, \code{x} \mapsto \Box, \code{y} \mapsto 2, \code{z} \mapsto \Box] & \longleftrightarrow &
    [\code{w} \mapsto v, \code{x} \mapsto \Box, \code{y} \mapsto 2, \code{z} \mapsto \Box]\\
{}  [\code{w} \mapsto v, \code{x} \mapsto \Box, \code{y} \mapsto \Box, \code{z} \mapsto \Box] & \longleftrightarrow &
    [\code{w} \mapsto v, \code{x} \mapsto \Box, \code{y} \mapsto \Box, \code{z} \mapsto \Box]
\end{eqnarray*}
where $\code{v} \in \{\Box,0\}$ so that each line above abbreviates two concrete
relationships; the lattice has the shape of a cube.  Because $w$ is not read or
written by \code{z := x + y}, it is preserved if present in the forward
direction or if required in the backward direction.  Because \code{z} is written
but not read, its initial value is always irrelevant.  To obtain the backward
slice of any other partial output, such as $[\code{w} \mapsto \Box, \code{x}
  \mapsto 1, \code{y} \mapsto \Box, \code{z} \mapsto 3]$, find the smallest
maximal partial output containing it, and take its backward slice, e.g.
$[\code{w} \mapsto \Box, \code{x} \mapsto 1, \code{y} \mapsto 1, \code{z}
  \mapsto \Box]$.

In the backward slicing rules for \ifKw instructions, we only reconstruct a
branch of the conditional that was actually taken during execution, leaving a
second branch as a \hole.  Importantly in these rules state $\mu_b$ is a minimal
state sufficient for an \ifKw condition to evaluate to a \True\ or
\False\ value.  That state is joined with state $\mu'$, which is a state
sufficient to evaluate the reconstructed branch of an \code{if}.

Rules for \whileKw slicing follow a similar approach.  It might seem that the
second rule for slicing \code{while}$_\truenkw$ is redundant because it is a
special case of the third \code{while}$_\truenkw$ rule if we allowed $c_w =
\hole$.  Indeed, that is the case on paper.  However, for the purpose of a
mechanised formalisation we require that these two rules are separate.  This
shows that formalising systems designed on paper can indeed be tricky and
require modifications tailored to solve mechanisation-specific issues.

Readers might have noticed that whenever a backward slicing rule from
Figure~\ref{fig:imp-bwd-cmd-slicing} returns \hole\ as an output program, the
state returned by the rule will be identical to the input state.  One could then
argue that we should reflect this in our rules by explicitly denoting that input
and output states are the same, e.g.
\begin{displaymath}
 \dfrac{\trace{T_1} \mu \searrow \mu, \Box}{\traceiftrue \mu \searrow \mu, \Box}
\end{displaymath}
\noindent
While it is true that in such a case states will be equal, this approach would
not be directly reflected in the implementation, where slicing is implemented as
a function and a result is always assigned to a new variable.  However, it would
be possible to prove a lemma about equality of input and output states for
\hole\ output programs, should we need this fact.

\subsection{An Extended Example of Backward Slicing}
\label{sec:slicing-extended-example}

We now turn to an extended example that combines all the programming constructs
of Imp\footnote{This example is adapted from~\cite{LecKosLeGt18}.}: assignments,
sequencing, conditionals, and loops.  Figure~\ref{fig:extended-example} shows a
program that divides integer \code{a} by \code{b}, and produces a quotient
\code{q}, remainder \code{r}, and result \code{res} that is set to $1$ if
\code{b} divides \code{a} and to $0$ otherwise.

\begin{figure}[h!]
\vspace{-0.25cm}
\begin{subfigure}[t]{0.45\textwidth}
\small
$[ \code{q} \mapsto 0, \code{r} \mapsto 0, \code{res} \mapsto 0, \code{a} \mapsto 4, \code{b} \mapsto 2 ]$
\begin{alltt}
r := a;
while ( b <= r ) do \{
  q := q + 1;
  r := r - b
\};
if ( ! (r = 0) )
then \{ res := 0 \}
else \{ res := 1 \}
\end{alltt}
$[ \code{q} \mapsto 2, \code{r} \mapsto 0, \code{res} \mapsto 1, \code{a} \mapsto 4, \code{b} \mapsto 2 ]$
\caption{Original program.}
\label{fig:extended-example-a}
\end{subfigure}
\hspace{1cm}
\begin{subfigure}[t]{0.5\textwidth}
\small
$[ \code{q} \mapsto \hole, \code{r} \mapsto \hole, \code{res} \mapsto \hole, \code{a} \mapsto 4, \code{b} \mapsto 2 ]$
\begin{alltt}
r := a;
while ( b <= r ) do \{
  \hole;
  r := r - b
\};
if ( ! (r = 0) )
then \{ \hole \}
else \{ res := 1 \}
\end{alltt}
$[ \code{q} \mapsto \hole, \code{r} \mapsto \hole, \code{res} \mapsto 1, \code{a} \mapsto \hole, \code{b} \mapsto \hole ]$
\caption{Backward slice w.r.t. $\code{res} \mapsto 1$.}
\label{fig:extended-example-b}
\end{subfigure}

 \caption{Slicing a program that computes whether $b$ divides $a$.}
 \label{fig:extended-example}
\end{figure}

To test whether $2$ divides $4$ we set $\code{a} \mapsto 4$, $\code{b}
\mapsto 2$ in the input state.  The remaining variables \code{q}, \code{r} and
\code{res} are initialised to $0$ (Figure~\ref{fig:extended-example-a}).  The
\whileKw loop body is executed twice; the loop condition is evaluated three times.
Once the loop has stopped, variable \code{q} is set to $2$ and variable \code{r}
to $0$.  Since the \ifKw condition is false we execute the \code{else} branch and
set \code{res} to $1$.  Figure~\ref{fig:extended-example-trace} shows the execution
trace.

\begin{figure}[h!]
\vspace{-0.25cm}
\begin{subfigure}[t]{0.42\textwidth}
\begin{equation*}
\begin{split}
(1)&\;\;\code{r} := \code{a}(4);\\
(2)&\;\;\code{while}_{\truenkw}\; (\code{b}(2) <= \code{r}(4))\; \code{do}\; \{\\
(3)&\;\;\;\; \code{q} := \code{q}(0) + 1;\;\code{r} := \code{r}(4) - \code{b}(2)\\
(4)&\;\;\};\;\\
(5)&\;\;\code{while}_{\truenkw}\; (\code{b}(2) <= \code{r}(2))\; \code{do}\; \{\\
(6)&\;\;\;\; \code{q} := \code{q}(1) + 1;\;\code{r} := \code{r}(2) - \code{b}(2)\\
(7)&\;\;\};\;\\
\end{split}
\end{equation*}
\end{subfigure}
\hspace{1cm}
\begin{subfigure}[t]{0.42\textwidth}
\begin{equation*}
\begin{split}
(8)&\;\;\code{while}_{\falsenkw}\; (\code{b}(2) <= \code{r}(0));\\
(9)&\;\;\code{if}_{\falsenkw}\; (\neg(\code{r}(0) = 0)) \;\code{else} \;\{\\
(10)&\;\;\;\;\code{res} := 1\\
(11)&\;\;\}
\end{split}
\end{equation*}
\end{subfigure}

 \caption{Trace of executing an example program for $\code{a} \mapsto 4$ and $\code{b} \mapsto 2$.}
 \label{fig:extended-example-trace}
\end{figure}

We now want to obtain an explanation of \code{res}.  We form a slicing criterion
by setting $\code{res} \mapsto 1$ (this is the value at the end of execution);
all other variables are set to \hole.

We begin by reconstructing the \ifKw conditional.  We apply the second rule for
$\code{if}_{\falsenkw}$ slicing (Figure~\ref{fig:extended-example-b}).  This is
because $c_2$, i.e. the body of this branch, backward slices to an assignment
\code{res := 1}, and not to a \hole\ (in which case the first rule for
$\code{if}_{\falsenkw}$ slicing would apply).  Assignment in the \code{else}
branch is reconstructed by applying the second rule for assignment slicing.
Since the value assigned to \code{res} is a constant it does not require
presence of any variables in the state.  Therefore state $\mu_a$ is empty.
Moreover, variable $\code{res}$ is erased in state $\mu$; joining of $\mu_a$ and
$\mu[\code{res} \mapsto \hole]$ results in an empty state, which indicates that
the code inside the \code{else} branch does not rely on the program state.
However, to reconstruct the condition of the \ifKw we need a state $\mu_b$ that
contains variable $\code{r}$.  From the trace we read that $\code{r} \mapsto 0$,
and so after reconstructing the conditional we have a state where $\code{r}
\mapsto 0$ and all other variables, including $\code{res}$, map to \hole.

We now apply the third rule for sequence slicing and proceed with reconstruction
of the \whileKw loop.  First we apply a trivial $\code{while}_{\falsenkw}$ rule.
The rule basically says that there is no need to reconstruct a \whileKw loop
that does not execute -- it might as well not be in a program.  Since the final
iteration of the \whileKw loop was reconstructed as a \hole, we reconstruct the
second iteration using the second $\code{while}_{\truekw}$ backward slicing
rule, i.e. the one where we have $\trace{T_w} \mu \searrow \mu_w, \hole$ as the
first premise.  We begin reconstruction of the body with the second assignment
$\code{r} := \code{r}(2) - \code{b}(2)$.  Recall that the current state assigns
$0$ to $\code{r}$.  The RHS is reconstructed using the second rule for backward
slicing of assignments we have already applied when reconstructing \code{else}
branch of the conditional.  An important difference here is that $\code{r}$
appears both in the LHS and RHS.  Reconstruction of RHS yields a state where
$\code{r} \mapsto 2$ and $\code{b} \mapsto 2$ (both values read from a trace),
whereas the current state contains $\code{r} \mapsto 0$.  Here it is crucial
that we erase $\code{r}$ in the current state before joining.  We apply third
rule of sequence slicing and proceed to reconstruct the assignment to $\code{q}$
using the first rule for assignment slicing (since $\code{q} \mapsto \hole$ in
the slicing criterion).  This reconstructs the assignment as a \hole.  We then
reconstruct the first iteration of the loop using the third
$\code{while}_{\truekw}$ slicing rule, since it is the case that $c_w \neq
\hole$.  Assignments inside the first iteration are reconstructed following the
same logic as in the second iteration, yielding a state where $\code{r} \mapsto
4$, $\code{b} \mapsto 2$, and other variables map to \hole.

Finally, we reconstruct the initial assignment \code{r := a}.  Since $\code{r}$
is present in the slicing criterion, we yet again apply the second rule for
assignment slicing, arriving at a partial input state $[ \code{q} \mapsto 0,
  \code{r} \mapsto 0, \code{res} \mapsto 0, \code{a} \mapsto 4, \code{b} \mapsto
  2 ]$ and a partial program shown in Figure~\ref{fig:extended-example-b}.

\section{Formalisation}
\label{sec:formalisation}

In the previous sections we defined the syntax and semantics of the Imp
language, and provided definitions of slicing in a Galois connection framework.
We have implemented all these definitions in the Coq proof assistant~\cite{Coq}
and proved their correctness as formal theorems.  The following subsections
outline the structure of our Coq development.  We provide references to the
source code by providing the name of file and theorem or definition as
\sref{filename.v}{theorem\_name, definition\_name}.  We will use \code{*} in
abbreviations like \code{*\_monotone} to point to several functions ending with
\code{\_monotone} suffix.  The whole formalisation is around 5.2k lines of Coq
code (not counting the comments).  Full code is available
online~\cite{the-code}.

\subsection{Lattices and Galois connections}
\label{sec:lattice-impl}

Our formalisation is built around a core set of definitions and theorems about
lattices and Galois connections.  Most importantly we define:

\begin{itemize}

 \item that a relation that is reflexive, antisymmetric and transitive is a
   partial order \sref{Lattice.v}{order}.  When we implement concrete
   definitions of ordering relations we require a proof that these
   implementations indeed have these three properties,
   e.g. \sref{ImpPartial.v}{order\_aexpPO}.

 \item what it means for a function $f : P \to Q$ to be monotone
   \sref{Lattice.v}{monotone}:
  \begin{equation*}
   \forall_{x, y}\;\, x \sqsubseteq_P y \implies f(x) \sqsubseteq_Q f(y)
  \end{equation*}

 \item consistency properties as given in Section~\ref{sec:slicing}
   \sref{Lattice.v}{inflation, def-\linebreak lation}.

 \item a Galois connection of two functions between lattices $P$ and $Q$ (see
   Definition~\ref{def:galois-connection} in Section~\ref{sec:slicing})
   \sref{Lattice.v}{galoisConnection}.
\end{itemize}
\noindent
We then prove that:

\begin{itemize}
  \item existence of a Galois connection between two functions implies their
    consistency and minimality \sref{Lattice.v}{gc\_implies\_consistency,
      gc\_implies\_\linebreak minimality}.

 \item two monotone functions with deflation and inflation properties form a
   Galois connection \sref{Lattice.v}{cons\_mono\_\_gc}.
\end{itemize}

Throughout the formalisation we operate on elements inside lattices of partial
expressions ($\prefix{a},\prefix{b}$, commands ($\prefix{c}$) or states
($\prefix{\mu}$).  We represent values in a lattice with an inductive data type
\code{prefix}\footnote{Name comes from a term ``prefix set'' introduced
  in~\cite{RicStoPerChe17} to refer to a set of all partial values smaller than
  a given value.  So a prefix set of a top element of a lattice denotes a set of
  all elements in that (complete) lattice.}  \sref{PrefixSets.v}{prefix} indexed
by the top element of the lattice and the ordering relation\footnote{In order to
  make the code snippets in the paper easier to read we omit the ordering
  relation when indexing \code{prefix}.}.  Values of \code{prefix} data type
store an element from a lattice together with the evidence that it is in the
ordering relation with the top element.  Similarly we define an inductive data
type \code{prefixO} \sref{PrefixSets.v}{prefixO} for representing ordering of
two elements from the same lattice.  This data type stores the said two elements
together with proofs that one is smaller than another and that both are smaller
than the top element of a lattice.

\subsection{Imp syntax and semantics}
\label{sec:imp-impl}

All definitions given in Figures~\ref{fig:imp-syntax}--\ref{fig:join-aexp} are
directly implemented in our Coq formalisation.

Syntax trees for Imp (Figure~\ref{fig:imp-syntax}), traces
(Figure~\ref{fig:imp-trace-syntax}) and partial Imp
(Figure~\ref{fig:partial-imp-syntax}) are defined as ordinary inductive data
types in \sref{Imp.v}{aexp, bexp, cmd}, \sref{Imp.v}{aexpT, bexpT, cmdT} and
\sref{ImpPartial.v}{aexpP, bexpP, cmdP}, respectively.  We also define functions
to convert Imp expressions to partial Imp expressions by rewriting from normal
syntax tree to a partial one \sref{ImpPartial.v}{aexpPartialize, bexpPartialize,
  cmdPartialize}.

Evaluation relations for Imp
(Figures~\ref{fig:imp-aexp-eval}--\ref{fig:imp-cmd-eval}) and ordering relations
for partial Imp (Figure~\ref{fig:partial-order-aexp}) are defined as inductive
data types with constructors indexed by elements in the relation
(\texttt{Imp.v}: \texttt{aevalR, bevalR, cevalR} and \texttt{ImpPartial.v}:
\texttt{aexpPO, bexpPO, comPO}), respectively.  For each ordering relation we
construct a proof of its reflexivity, transitivity, and antisymmetry, which
together proves that a given relation is a partial order
\sref{ImpPartial.v}{order\_aexpPO, order\_\linebreak bexpPO, order\_comPO}.

Join operations (Figure~\ref{fig:join-aexp}) are implemented as functions
\sref{ImpPartial.v}{aexpLUB, bexpLUB, comLUB}.  Their implementation is
particularly tricky.  Coq requires that all functions are total.  We know that
for two elements from the same lattice a join always exists, and so a join
function is a total one.  However, we must guarantee that a join function only
takes as arguments elements from the same lattice.  To this end a function takes
three arguments: top element $e$ of a lattice and two \code{prefix} values
$e_1$, $e_2$ indexed by the top element $e$.  So for example if $e$ is a
variable name $x$, we know that each of $e_1$ and $e_2$ is either also a
variable name $x$ or a \hole.  However, Coq does not have a built-in dependent
pattern match and this leads to complications.  In our example above, even if we
know that $e$ is a variable name $x$ we still have to consider cases for $e_1$
and $e_2$ being a constant or an arithmetic operator.  These cases are of course
impossible, but it is the programmer's responsibility to dismiss them
explicitly.  This causes further complications when we prove lemmas about
properties of join, e.g.:
\begin{displaymath}
e_1 \sqleq e \wedge e_2 \sqleq e \implies (e_1 \join e_2) \sqleq e
\end{displaymath}
\noindent
This proof is done by induction on the top element of a lattice, where $e$,
$e_1$, and $e_2$ are all smaller than that element.  The top element limits the
possible values of $e$, $e_1$, and $e_2$ but we still have to consider the
impossible cases and dismiss them explicitly.

\subsection{Program state}
\label{sec:state-impl}

Imp programs operate by side-effecting on a program state.  Handling of the
state was one of the most tedious parts of the formalisation.

State is defined as a data type isomorphic to an association list that maps
variables to natural number values \sref{ImpState.v}{state}.  Partial state is
defined in the same way, except that it permits partial values, i.e. variables
can be mapped to a hole or a numeric value \sref{ImpState.v}{stateP}.  We assume
that no key appears more than once in a partial state.  This is not enforced in
the definition of \code{stateP} itself, but rather defined as a separate
inductive predicate \sref{ImpState.v}{statePWellFormed} that is explicitly
passed as an assumption to any theorem that needs it.  We also have a
\code{statePartialize} function that turns a state into a partial state.  This
only changes representation from one data type to another, with no change in the
state contents.

For partial states we define an ordering relation as a component-wise ordering
of partial values inside a state \sref{ImpState.v}{statePO}.  This assumes that
domains of states in ordering relation are identical (same elements in the same
order), which allows us to formulate lemmas such as:
\begin{displaymath}
[\,] \leq \mu \implies \mu = [\,]
\end{displaymath}
\noindent
This lemma says that if a partial state $\mu$ is larger than an state with empty
domain then $\mu$ itself must have an empty domain.

We also define a join operation on partial states, which operates element-wise
on two partial states from the same lattice \sref{ImpState.v}{stateLUB}.

As already mentioned in Section~\ref{sec:imp}, the domain of the state is fixed
throughout the execution.  This means that updating a variable that does not
exist in a state is a no-op, i.e. it returns the original state without any
modifications.  This behaviour is required to allow comparison of states before
and after an update.  Consider this lemma:
\begin{displaymath}
(\mu_1 \leq \mu_2) \wedge (v_1 \leq \mu_2(k)) \implies \mu_1[k \mapsto v_1] \leq
  \mu_2
\end{displaymath}
\noindent
It says that if a partial state $\mu_1$ is smaller than $\mu_2$ and the value
stored in state $\mu_2$ under key $k$ is greater than $v_1$ then we can assign
$v_1$ to key $k$ in $\mu_1$ and the ordering between states will be maintained.
A corner-case for this theorem is when the key $k$ is absent from the states
$\mu_1$ and $\mu_2$.  Looking up a non-existing key in a partial state returns a
\hole.  If $k$ did not exist in $\mu_2$ (and thus $\mu_1$ as well) then
$\mu_2(k)$ would return \hole\ and so $v_1$ could only be a \hole\ (per second
assumption of the theorem).  However, if we defined semantics of update to
insert a non-existing key into the state, rather than be a no-op, the conclusion
of the theorem would not hold because domain of $\mu_1[k \mapsto v_1]$ would
contain $k$ and domain of $\mu_2$ would not, thus making it impossible to define
the ordering between the two states.

The approach described above is one of several possible design choices.  One
alternative approach would be to require evidence that the key being updated
exists in the state, making it impossible to attempt update of non-existent
keys.  We have experimented with this approach but found explicit handling of
evidence that a key is present in a state very tedious and seriously
complicating many of the proofs.  In the end we decided for the approach
outlined above, as it allowed us to prove all the required lemmas, with only
some of them relying on an explicit assumption that a key is present in a state.
An example of such a lemma is:
\begin{displaymath}
\mu[k \mapsto v](k) = v
\end{displaymath}
\noindent
which says that if we update key $k$ in a partial state $\mu$ with value $v$ and
then immediately lookup $k$ in the updated state we will get back the $v$ value
we just wrote to $\mu$.  However, this statement only holds if $k$ is present in
$\mu$.  If it was absent the update would return $\mu$ without any changes and
then lookup would return \hole, which makes the theorem statement false.  Thus
this theorem requires passing explicit evidence that $k \in \mathsf{dom}(\mu)$
in order for the conclusion to hold.

Formalising program state was a tedious task, requiring us to prove over sixty
lemmas about the properties of operations on state, totalling over 800 lines of
code.

\subsection{Slicing functions}
\label{sec:slicing-impl}

Slicing functions implement rules given in
Figures~\ref{fig:imp-fwd-aexp-slicing}--\ref{fig:imp-bwd-cmd-slicing}.  We have
three separate forward slicing functions, one for arithmetic expressions, one
for logical expressions and one for imperative commands
\sref{ImpSlicing.v}{aexpFwd, bexpFwd, comFwd}.  Similarly for backward slicing
\sref{ImpSlicing.v}{aexpBwd, bexpBwd, comBwd}.

In Section~\ref{sec:slicing} we said that forward and backward slicing functions
operate between two lattices.  If we have an input $p$ (an arithmetic or logical
expression or a command) with initial state $\mu$ that evaluates to output $p'$
(a number, a boolean, a state) and records a trace $T$ then \fwdFt{T}{(p,\mu)}
is a forward slicing function parametrized by $T$ that takes values from lattice
generated by $(p,\mu)$ to values in lattice generated by $p'$.  Similarly
\bwdFt{T}{p'} is a backward slicing function parametrized by $T$ that takes
values from lattice generated by $p'$ to values in lattice generated by
$(p,\mu)$.  Therefore our implementation of forward and backward slicing
functions has to enforce that:

\begin{enumerate}
 \item\label{enum:slicing-cond-1} $(p, \mu)$ evaluates to $p'$ and records trace
   $T$
 \item\label{enum:slicing-cond-2} forward slicing function takes values from
   lattice generated by $(p,\mu)$ to lattice generated by $p'$
 \item\label{enum:slicing-cond-3} backward slicing function takes values from
   lattice generated by $p'$ to lattice generated by $(p,\mu)$
\end{enumerate}

To enforce the first condition we require that each slicing function is
parametrized by inductive evidence that a given input $(p, \mu)$ evaluates to
$p'$ and records trace $T$.  We then define input and output types of such
slicing functions as belonging to relevant lattices, which is achieved using the
\code{prefix} data type described in Section~\ref{sec:lattice-impl}.  This
enforces the conditions above.  For example, the type signature of the forward
slicing function for arithmetic expressions looks like this:

\small
\begin{snippet}
Fixpoint aexpFwd {st : state} {a : aexp}
   {v : nat} {t : aexpT}
   (ev : t :: a, st \\ v):
   (prefix a * prefix st) -> prefix v.
\end{snippet}
\normalsize
\noindent
Here \code{ev} is evidence that arithmetic expression \code{a} with input state
\code{st} evaluates to a natural number \code{v} and records an execution trace
\code{t}.  The \code{t :: a, st \bslash\bslash\, v} syntax is a notation for the
evaluation relation.  The first four arguments to \code{aexpFwd} are in curly
braces, denoting they are implicit and can be inferred from the type of
\code{ev}.  The function then takes values from the lattice generated by
\code{(a, st)} and returns values in the lattice generated by \code{v}.

In the body of a slicing function we first decompose the evaluation evidence
with pattern matching.  In each branch we implement logic corresponding to
relevant slicing rules defined in
Figures~\ref{fig:imp-fwd-aexp-slicing}--\ref{fig:imp-bwd-cmd-slicing}.  Premises
appearing in the rules are turned into recursive calls.  If necessary, results
of these calls are analysed to decide which rule should apply.  For example,
when backward slicing sequences we analyse whether the recursive calls return
holes or expressions to decide which of the rules should apply.

The implementation of the slicing functions faces similar problems as the
implementation of joins described in Section~\ref{sec:imp-impl}.  When we
pattern match on the evaluation evidence, in each branch we are restricted to
concrete values of the expression being evaluated.  For example, if the last
step in the evaluation was an addition, then we know the slicing criterion is a
partial expression from a lattice formed by expression $a_1 + a_2$.  Yet we have
to consider the impossible cases, e.g. having an expression that is a constant,
and dismiss them explicitly.  Moreover, operating inside a lattice requires us
not to simply return a result, but also provide a proof that this result is
inside the required lattice.  We rely on Coq's \code{refine} tactic to construct
the required proof terms.  All of this makes the definitions of slicing
functions very verbose.  For example, forward slicing of arithmetic expressions
requires over 80 lines of code with over 60 lines of additional boilerplate
lemmas to dismiss the impossible cases.

For each slicing function we state and prove a theorem that it is monotone
\sref{ImpSlicing.v}{*\_monotone}.  For each pair of forward and backward slicing
functions we state theorems that these pairs of functions have deflation and
inflation properties \sref{ImpSlicing.v}{*\_deflating, *\_inflating}, as defined
in Section~\ref{sec:slicing}.  Once these theorems are proven we create
instances of a general theorem \code{cons\_mono\_\_gc}, described in
Section~\ref{sec:lattice-impl}, which proves that our definitions form a Galois
connection and are thus a correctly defined pair of slicing functions.  We also
create instances of the \code{gc\_implies\_minimality} theorem, one instance for
each slicing function.  This provides us with a formalisation of all the
correctness properties, proving our main result:
\begin{theorem}
  Suppose $\mu_1, c \Rightarrow \trace{T} \mu_2$.  Then there exist total,
  monotone functions $\fwdFt{T}{(c,\mu_1)} : \prefix{c} \times \prefix{\mu_1}
  \to \prefix{\mu_2}$ and $\bwdFt{T}{\mu_2} : \prefix{\mu_2} \to \prefix{c}
  \times \prefix{\mu_1}$. Moreover, $\bwdFt{T}{\mu_2} \dashv
  \fwdFt{T}{(c,\mu_1)}$ form a Galois connection and in particular satisfy the
  minimality, inflation (consistency), and deflation properties.
\end{theorem}
Here, the forward and backward slicing judgements are implemented as
functions $\fwdFt{T}{(c,\mu_1)}$ and $\bwdFt{T}{\mu_2}$.

\section{Related and Future Work}
\label{sec:related-work}

During the past few decades a plethora of slicing algorithms has been presented
in the literature.  See \cite{Silva2008AnAO} for a good, although now slightly
out of date, survey.  Most of these algorithms have been analysed in a formal
setting of some sort using pen and paper.  However, work on formalising slicing
in a machine checked way has been scarce.  One example of such a development is
\cite{PereraGC16}, which formalises dynamic slicing for $\pi$-calculus in Agda
using a Galois connection framework identical to the one used in this paper.
The high-level outline of the formalisation is thus similar to ours.  However,
details differ substantially, since \cite{PereraGC16} formalises a completely
different slicing algorithm for concurrent processes using a different proof
assistant.  Another example of formalising slicing in a proof assistant is
\cite{Blazy15}, where Coq is used to perform an \emph{a posteriori} validation
of a slice obtained using an unverified program slicer.  This differs from our
approach of verifying correctness of a slicing algorithm itself.  We see our
approach of verifying correctness of the whole algorithm as a significant
improvement over the validation approach.  In a more recent work L{\'e}chenet et
al.~\cite{LecKosLeGt18} introduce a variant of static slicing known as relaxed
slicing and use Coq to formalise the slicing algorithm.  Their work is identical
in spirit to ours and focuses on the Imp language\footnote{Authors
  of~\cite{LecKosLeGt18} use the name WHILE, but the language is the same.} with
an extra \code{assert} statement.

Galois connections have been investigated previously as a tool in the
mathematics of program construction, for example by
Backhouse~\cite{backhouse00acmpc} and more recently by Mu and
Oliveira~\cite{mu12jlamp}.  As discussed in Section~\ref{sec:intro}, Galois
connections capture a common pattern in which one first specifies a space of
possible solutions to a problem, the ``easy'' part, via one adjoint, and defines
the mapping from problem instances to optimal solutions, the ``hard'' part, as
the Galois dual.  In the case of slicing, we have used the goal of obtaining a
verifiable Galois connection, along with intuition, to motivate choices in the
design of the forward semantics, and it has turned out to be easier for our
correctness proof to define both directions directly.

Mechanised proofs of correctness of calculational reasoning has been considered
in the Algebra of Programming in Agda (AOPA) system~\cite{mu09jfp}, and
subsequently extended to include derivation of greedy algorithms using Galois
connections~\cite{chiang16jlamp}.  Another interesting, complementary approach
to program comprehension is Gibbons' \emph{program fission}~\cite{gibbons06mpc},
in which the fusion law is applied ``in reverse'' to an optimized, existing
program in order to attempt to discover a rationale for its behavior: for
example by decomposing an optimized word-counting program into a ``reforested''
version that decomposes its behavior into ``construct a list of all the words''
and ``take the length of the list''.  We conjecture that the traces that seem to
arise as a natural intermediate structure in program slicing might be viewed as
an extreme example of fission.

An important line of work on slicing theory focuses on formalising different
slicing algorithms within a unified theoretical framework of \emph{program
  projection}~\cite{Binkley06}.  Authors of that approach develop a precise
definition of what it means that one form of slicing is weaker than another.
However, our dynamic slicing algorithm does not fit the framework as presented
in~\cite{Binkley06}.  We believe that it should be possible to extend the
program projection framework so that it can encompass slicing based on Galois
connections but this is left as future work.

\section{Summary}

Program slicing is an important tool for aiding software development.  It is
useful when creating new programs as well as maintaining existing ones.  In this
paper we have developed and formalised an algorithm for dynamic slicing of
imperative programs.  Our work extends the line of research on slicing based on
the Galois connection framework.  In the presented approach slicing consists of
two components: forward slicing, that allows to execute partial programs, and
backward slicing, that allows to ``rewind'' program execution to explain the
output.

Studying slicing in a formal setting ensures the reliability of this technique.
We have formalised all of the theory presented in this paper using the Coq proof
assistant.  Most importantly, we have shown that our slicing algorithms form a
Galois connection, and thus have the crucial properties of consistency and
minimality.  One interesting challenge in our mechanisation of the proofs was
the need to modify some of the theoretical developments so that they are easier
to formalise in a proof assistant -- c.f. overlapping rules for backward slicing
of while loops described in Section~\ref{sec:imp-backward-slicing}.

Our focus in this paper was on a simple programming language Imp.  This work
should be seen as a stepping stone towards more complicated formalisations of
languages with features like (higher-order) functions, arrays, and pointers.
Though previous work~\cite{RicStoPerChe17} has investigated slicing based on
Galois connections for functional programs with imperative features, our
experience formalising slicing for the much simpler Imp language suggests that
formalising a full-scale language would be a considerable effort. We leave this
as future work.

\section*{Acknowledgements}
We gratefully acknowledge help received from Wilmer Ricciotti during our work on
the Coq formalisation, and Jeremy Gibbons for comments on a draft.  This work
was supported by ERC Consolidator Grant Skye (grant number 682315).

\balance
\bibliographystyle{splncs04}
\bibliography{bibliography}

\end{document}